\documentclass[preprint,showpacs]{revtex4}
\usepackage{graphicx}
\begin{document}
\title{Scaling in the crossover from random to correlated growth}
\author{F. D. A. Aar\~ao Reis}
\affiliation{
Instituto de F\'\i sica, Universidade Federal Fluminense,\\
Avenida Litor\^anea s/n, 24210-340 Niter\'oi RJ, Brazil}
\date{\today}

\begin{abstract}

In systems where deposition rates are high compared to diffusion, desorption and
other mechanisms that generate correlations, a crossover from random
to correlated growth of surface roughness is expected at a characteristic time
$t_0$. This crossover is analyzed in lattice models via scaling arguments, with
support from simulation results presented here and in other authors works.
We argue that the amplitudes of the saturation roughness and of the saturation
time $t_\times$ scale as ${t_0}^{1/2}$ and $t_0$, respectively.
For models with lateral aggregation, which typically are in the
Kardar-Parisi-Zhang
(KPZ) class, we show that $t_0\sim p^{-1}$, where $p$ is the probability of the
correlated aggregation mechanism to take place. However, $t_0\sim p^{-2}$ is
obtained in solid-on-solid models with single particle deposition attempts. 
This group includes models in various universality classes, with numerical
examples being provided in the Edwards-Wilkinson (EW), KPZ and Villain-Lai-Das
Sarma (nonlinear molecular-beam epitaxy) classes. Most applications are for
two-component models in which random deposition, with probability $1-p$,
competes with a correlated aggregation process with probability $p$. However,
our approach can be extended to other systems
with the same crossover, such as the generalized restricted solid-on-solid
model with maximum height difference $S$, for large $S$. Moreover, the scaling
approach applies to all dimensions. In the particular case of one-dimensional
KPZ
processes with this crossover, we show that
$t_0\sim\nu^{-1}$ and $\nu\sim \lambda^{2/3}$, where $\nu$ and $\lambda$ are
the coefficients of the linear and nonlinear terms of the associated KPZ
equations. The applicability of previous results on models in the EW and
KPZ classes is discussed.

\end{abstract}

\maketitle

\section{Introduction}
\label{intro}

The large number of technological applications of thin films and multilayers
motivates the study of continuous and atomistic
models for the growth of those structures \cite{mbe,barabasi,krug}. Their
morphology is often the product of a competition between different growth
dynamics, thus theoretical models representing such
features received increasing attention in recent years. Some examples are
the models with aggregation of different species of
particles \cite{cerdeira,poison1,kotrla} or those mixing different
microscopic aggregation
rules for the same species
\cite{albano1,shapir,julien,albano2,albano3,tales,chamereis,
lidia,kolakowska1,lam,kolakowska2}.

In processes starting from a flat surface where the deposition rate is high
compared to ad-particle diffusion coefficients, a random growth is initially
observed, with negligible spatial correlations in the local thicknesses.
Subsequently, diffusion, desorption
and other mechanisms introduce surface correlations and, consequently, a
crossover from random to correlated growth is observed.
The simplest quantitative characteristic of the film surface which reveals
this crossover is the average roughness (or interface width), defined as the rms
fluctuation of the height $h$ around its average position $\overline{h}$:
$W(L,t)\equiv
{\left[ { \left<  {\left( h - \overline{h}\right) }^2  \right> } \right]
}^{1/2}$. The aim of this work is to study scaling relations for the surface
roughness in lattice models which present this crossover.

In a random, uncorrelated growth, the roughness increases as
\begin{equation}
W_r\sim t^{1/2} .
\label{wrandom}
\end{equation}
On the other hand, in correlated growth processes, it
is expected to obey the Family-Vicsek scaling relation \cite{fv}
\begin{equation}
W(L,t) \approx A L^{\alpha} f\left( \frac{t}{t_\times}\right) ,
\label{fv}
\end{equation}
where $L$ is the system size, $\alpha$ is the roughness exponent, $A$ is a
model-dependent constant, $f$
is a scaling function such that $f\sim 1$ in the regime of roughness saturation
($t\to\infty$) and $t_\times$ is the characteristic time of crossover to
saturation. $t_\times$ scales with the system size as 
\begin{equation}
t_\times \approx BL^z , 
\label{scalingtau}
\end{equation}
where $z$ is the dynamic exponent and $B$ is another model-dependent
amplitude. For $t\ll t_\times$ (after a possible crossover to correlated
growth), the roughness scales as
\begin{equation}
W\approx Ct^\beta , 
\label{scalingwgr}
\end{equation}
where $C$ is constant and $\beta=\alpha/z$ is the growth exponent. In this
growth regime, $f(x)\sim x^\beta$ in Eq. (\ref{fv}).

Fig. 1 illustrates the time evolution of the surface roughness in
systems which present a crossover from random to correlated growth at a
characteristic time $t_0$ (this time is also called $t_{\times 1}$ and $\tau_h$
by other authors \cite{albano3,lam,chien}). For $t<t_0$, $W$ increases as Eq.
(\ref{wrandom}), and the
reduced slope in the subsequent regime ($t_0\ll t\ll t_\times$) is a signature
of the smoothing effect of correlations. In systems where a correlation
mechanism and a random growth mechanism are simultaneously present, a different
balance of these mechanisms changes the crossover time $t_0$, which affects
the amplitudes $A$, $B$ and $C$ (Eqs. \ref{fv}, \ref{scalingtau} and
\ref{scalingwgr}).

This was observed by Albano and co-workers \cite{albano1,albano2,albano3}, who
studied numerically two-component models in which 
different rules for the aggregation of the same species are chosen, with
complementary probabilities (they are simply called competitive models by
several authors
\cite{albano1,albano2,albano3,tales,chamereis,lidia,lam}). 
The first component of those models, with probability $1-p$, was random
deposition (RD): particles attach to the top of the column of
incidence independently of the neighboring heights. In their first competitive
model (called RD-BD model), the second component was ballistic deposition (BD)
\cite{vold}, with probability $p$. In their second competitive
model (called RD-RDSR), the correlated component was random deposition with
surface relaxation (RDSR) \cite{family}.
As expected from the absence of correlations in RD, the universal scaling
exponents $\alpha$, $\beta$ and $z$ of the competitive models are those of the
universality class of the correlated component \protect\cite{kolakowska2}.
However, the scaling amplitudes are
affected, and scale with $p$ as \cite{albano1,albano2,albano3}:
\begin{equation}
A\sim p^{-\delta} ,
\label{defdelta}
\end{equation}
\begin{equation}
B\sim p^{-y} ,
\label{defy}
\end{equation}
and 
\begin{equation}
C\sim p^{-\gamma} .
\label{defgamma}
\end{equation}
For the RD-BD model, $\delta =1/2$
and $y=1$ were obtained in all substrate dimensions $d$ \cite{albano2,albano3}.
For the RD-RDSR model, $\delta =1$ and $y=2$ were obtained
\cite{albano1,albano3}. In both competitive models, the exponent $\gamma$
depended on $d$.

In a recent work, Braunstein and Lam \cite{lam} explained the differences
between those systems through scaling arguments which account for the average
height increase and fluctuations during the time interval between two
correlated deposition events. Previously, the equation representing the RD-RDSR
model in the continuum limit was also derived \cite{lidia}.
Since those works analyzed particular models in the Edwards-Wilkinson (EW)
\cite{ew}
and in the Kardar-Parisi-Zhang (KPZ) \cite{kpz} classes,
the picture that emerged from those results was that the
exponents $\delta$ and $y$ are related to the universality class of
the dominant process: $\delta =1/2$ and $y=1$ in the crossover to KPZ, $\delta
=1$ and $y=2$ in the crossover to EW. The dominant behavior of another
competitive model in Ref. \protect\cite{lam} was not trivial to infer, but the
crossover exponents to KPZ and EW were the same. However, recent numerical work
by Kolakowska et al \cite{kolakowska1,kolakowska2} show that $\delta =1$ and
$y=2$ is also found in two-component models in the KPZ class. Thus, a complete
theoretical explanation of the values of these crossover exponents is still
lacking.

Here we will present a scaling approach which provides such explanation
through a connection of roughness amplitudes, crossover times and the
microscopic rules of the
lattice models. It can be applied to all spatial dimensions and to all types of
correlated growth, and agrees with our numerical results for models
in three different universality classes and with numerical results from other
authors works \cite{albano2,albano3,chien}.

We will argue that $A\sim {t_0}^{1/2}$ and $B\sim t_0$ in the models with random
to correlated crossover, so that $t_0$ acts as a time dilatation factor.
The relations between $t_0$ and the parameters of the
discrete models are subsequently obtained by scaling arguments, partially
rephrasing those of Braunstein and Lam \cite{lam}. These arguments lead to a
separation of
the lattice models in two groups: the first one includes solid-on-solid models
with single particle deposition attempts, for which $\delta =1$ and $y=2$, and
the second one includes models with lateral aggregation, for which $\delta
=1/2$ and $y=1$. This classification is independent of the universality class
of the correlated process, consequently KPZ processes may be found in both
groups, i. e. with different crossover exponents. Moreover, our theoretical
approach comprises not only two-component models involving RD, but can be
extended to other models with the same crossover.
One example is the generalized restricted solid-on-solid (RSOS) model, with
maximum height difference $S$ between the neighbors \cite{chien}, which presents
that crossover for large $S$ ($S^{-1}$ substitutes $p$ in Eqs. \ref{defdelta},
\ref{defy} and \ref{defgamma}) and which will be studied numerically here.

The theoretical analysis is motivated in Sec. II, with the presentation of
several discrete models showing the random to correlated crossover and the
discussion of their simulation results. The scaling approach is presented in
Sec. III, and in Sec. IV we summarize our conclusions.

\section{Lattice models and simulation results}

First we recall the models with crossover from random to correlated growth
previously studied by other authors.

The first one is RD-BD \cite{albano2}. In pure BD
($p=1$), the
particle sticks at first contact with a nearest neighbor, as illustrated in
Fig. 2a, which leads to the formation of a porous deposit.
Simulations of the RD-BD model gave $\delta\approx 1/2$ and $y\approx 1$ in
$d=1$, $d=2$ and $d=3$ \cite{albano2,albano3}, while $\gamma$ depended on $d$.
From the Family-Vicsek relation (\ref{fv}) and the $p$-scaling for the
amplitudes $A$, $B$ and $C$ (Eqs. \ref{defdelta}, \ref{defy} and
\ref{defgamma}), Albano and co-workers proposed that
\begin{equation}
\beta y-\delta+\gamma = 0
\label{scalingalbano}
\end{equation}
in any $d$, which agrees with their simulation results \cite{albano2,albano3}.
Here $\beta$ is the growth exponent of the KPZ class.

The KPZ equation, which describes BD in the continuum limit, is
\begin{equation}
{{\partial h}\over{\partial t}} = \nu{\nabla}^2 h + {\lambda\over 2}
{\left( \nabla h\right) }^2 + \eta (\vec{x},t) ,
\label{kpz}
\end{equation}
where $h$ is the height at the position $\vec{x}$ in a
$d$-dimensional substrate at time $t$, $\nu$ represents a surface tension,
$\lambda$ represents the excess velocity and $\eta$ is a Gaussian
noise~\cite{barabasi,kpz} with zero mean and co-variance $\langle
\eta\left(\vec{x},t\right) \eta (\vec{x'},t')\rangle = D\delta^d
(\vec{x}-\vec{x'} ) \delta\left( t-t'\right)$.
In $d=1$, the exact KPZ exponents are $\alpha=1/2$, $\beta=1/3$
and $z=3/2$, and in $d\geq 2$ approximate values are given in Refs.
\protect\cite{marinari} and \protect\cite{kpz2d}. In models with a crossover
from random to KPZ growth, small values of $\nu$ and $\lambda$ are expected in
the corresponding KPZ equation.

The second model studied by those authors was RD-RDSR \cite{albano1}.
In pure RDSR ($p=1$), the incident particle diffuses to the column with
minimum height in its nearest neighborhood \cite{family}. RDSR is described by
the EW
equation, which corresponds to Eq. \ref{kpz} with $\lambda =0$. In $d=1$,
$d=2$ and $d=3$, the exponents $\delta=1$ and $y=2$ \cite{albano3} were
obtained for the RD-RDSR model, while the exponent $\gamma$ also depended on
$d$.

Other models involving competition with RD in $d=1$ were recently proposed in
Refs. \protect\cite{kolakowska1} and \protect\cite{kolakowska2}. One example is
a model whose correlated component allows aggregation of the
incident particle only at surface minima \cite{kolakowska2}. It is in the KPZ
class, but has $\delta\approx 1$ and $y\approx 2$.

In the following, we will present our numerical results for three models
devised to broaden the investigation on the crossover exponents.

The first one is also a two-component model, involving RD with probability
$1-p$. With probability $p$, the aggregation is possible only if the height of
the column of incidence does not exceed the heights of any of its neighbors,
otherwise the aggregation attempt is rejected, as shown in Fig. 2b. In other
words, aggregation is possible only in valleys or plateaus.
This model mimics a competition between RD and RSOS deposition, thus we call it
the RD-RSOS model. We refer to the correlated component as RSOS because it
works against the formation of large local slopes, as illustrated in Fig. 2b
(there, the attempts at columns $3$ and $9$ are rejected because one or more
neighbors have smaller heights). In the pure RSOS model \cite{kk} ($p=1$), the
above aggregation rule implies ${\Delta h}_{max}=1$ between neighboring
columns.

The RD-RSOS model was simulated in one-dimensional
lattices with $32\leq L\leq 512$ for
some values of $p$ in the range $0.12\leq p\leq 0.4$. ${10}^4$ realizations
were obtained for the smallest lattices and ${10}^3$ for the largest ones.
For the same values of $p$, the model was simulated in substrates
with $L=8192$ during the random and the KPZ growth regimes.

In Fig. 3 we show the time evolution of the roughness for $p=0.15$ and $p=0.3$
in lattices with $L=256$.
For both values of $p$, it is clear that the roughness behaves as in the sketch
of Fig. 1.
The saturation regimes are clearer in the linear plot of the inset, which
shows that the saturation roughness approximately doubles when $p$ is reduced
from $0.3$ to $0.15$. Using Eqs. \ref{fv} and \ref{defdelta}, this suggests
$\delta \approx 1$. The parallel lines of the KPZ growth regimes illustrate the
behavior described by Eqs. (\ref{scalingwgr}) and (\ref{defgamma}), with the
amplitude $C$ increasing as $p$ decreases.

In order to obtain reliable estimates of the crossover exponents, we analyzed
the effects of finite $L$, $t$ and $p$, taking the limits $L\to\infty$,
$t\to\infty$, $p\to 0$ when appropriate. This procedure was proved to be
essential to avoid erroneous conclusions on the class of several growth models
(see, e. g. the analysis of BD in \protect\cite{balfdaar}).
For a fixed value of $p$, our first step is to extrapolate $W_{sat}/L^\alpha$ to
$L\to\infty$, using $\alpha=1/2$, as illustrated in Fig. 4a for $p=0.25$. From
Eq. (\ref{fv}), the asymptotic value of that ratio is the amplitude $A(p)$
($A\approx 3.18$ in Fig. 4a). For each
pair of subsequent probabilities $p'$ and $p''$, we define effective exponents
\begin{equation}
\delta_p = \frac{\ln{\left[ A{\left( p'\right)} / A{\left( p''\right)}
\right]}} {\ln{p''/p'}} 
\label {defdeltaef}
\end{equation}
where $p$ is an average probability:
\begin{equation}
p\equiv \sqrt{\left( p'p''\right)} .
\label {defp}
\end{equation}
As $p'$ and $p''$ decrease, $p\to 0$ and $\delta_p\to \delta$.
In Fig. 4b we show $\delta_p$ versus $p^2$, which suggests $\delta\approx 1$. 

The first step to estimate $\gamma$ is to extrapolate $W/t^{1/3}$ in the growth
regime, using the data from lattices with $L=8192$. For fixed $p$, that ratio
converges to $C(p)$ as
$t\to\infty$ (Eq. \ref{scalingwgr}). This is illustrated in Fig. 4c for
$p=0.2$, where we obtain $C\approx 1.22$ (this extrapolation
represents the long time behavior in an infinitely large substrate). Effective
exponents $\gamma_p$ were calculated
from $C(p)$ along the same lines of the calculation of $\delta_p$ in Eq.
(\ref{defdeltaef}). They are shown in Fig. 4d as a function of $p^2$, which
suggests $\gamma\approx 1/3$ as $p\to 0$. The estimates of $\delta$ and $\gamma$
and Eq. (\ref{scalingalbano}) give $y\approx 2$ for the RD-RSOS model.

Thus, although the RD-RSOS model is in the KPZ class, similarly to the
RD-BD model, its crossover exponents $\delta$ and $y$ are the same of the
RD-RDSR model, which is in the EW class.

The second model analyzed here does not involve competition of aggregation
rules. It is called generalized RSOS
model, and was
originally proposed in Refs. \protect\cite{kk} and \protect\cite{alanissila}.
The incident particle can aggregate at a certain column only if the
height differences between neighboring columns do not exceed an integer value
$S$, otherwise the deposition attempt is rejected. The version with $S=3$ is
illustrated in Fig. 2c.
For large $S$, with an initially flat substrate, random growth occurs
until a significant fraction of neighboring columns has height difference $S$.
Subsequently, KPZ growth is observed due to the rejection of
deposition attempts.

The generalized RSOS model was studied numerically by Chien et
al~\cite{chien}, who obtained $t_0\sim S^{2.06}$ for large $S$, in
agreement with their scaling arguments, which give $t_0\sim S^2$. However, the
roughness amplitudes were not calculated there.

Here it was simulated until saturation in one-dimensional
lattices with $32\leq L\leq 512$, for several values of $S$ in the range
$4\leq S\leq 32$. The number of realizations was ${10}^4$ for the smallest
lattices and ${10}^3$ for the largest ones. We also simulated the model in
$L=8192$ up to $t\gg t_x$, with ${10}^3$ realizations for each $S$.

For any $S$, the height difference ${\Delta h}\equiv h_{i+1}-h_i$ of
neighboring columns can assume $2S+1$ different values. However, deposition
attempts are rejected only when $\Delta h$ is $-S$ or $+S$, because those
attempts would lead to height differences $-(S+1)$ and $S+1$. For large $S$, it
is reasonable to assume that all values of $\Delta h$ have nearly the same
probability, thus the probability of rejecting aggregation is of order
$\frac{2}{2S+1}\approx S^{-1}$. Since aggregation rejection is the mechanism
to spread correlations, $S^{-1}$ plays the same role of $p$ in the other
competitive models. Thus we assume that
\begin{equation}
A(S) \sim S^\delta ,
\label{scalingAS}
\end{equation}
\begin{equation}
B(S) \sim S^y ,
\label{scalingBS}
\end{equation}
and
\begin{equation}
C(S) \sim S^\gamma 
\label{scalingCS}
\end{equation}
in the generalized RSOS model.

Estimates of the crossover exponents $\delta$ and $\gamma$ were obtained along
the same lines of the RD-RSOS model described above.
First, for fixed $S$, the extrapolation of $W_{sat}/L^{1/2}$ provided estimates
of the amplitude $A(S)$. Finite-size estimates of the exponent $\delta$ are
given by
\begin{equation}
\delta_S = \frac{\ln{\left[ A{\left( S\right)} / A{\left(
S/2\right)} \right]}} {\ln{2}} .
\label{defdeltaefS}
\end{equation}
Their values are shown in Fig. 5a as a function of $1/S$, suggesting
$\delta\approx 1$ asymptotically.
Amplitudes $C(S)$ were obtained from the extrapolation of $W/t^{1/3}$ in the
KPZ growth regime of large substrates, and effective exponents $\gamma_S$ were
defined analogously to Eq. (\ref{defdeltaefS}). They are shown in Fig. 5b as a
function of $1/S$, suggesting $\gamma\approx 1/3$ asymptotically. Those values
also lead to $y\approx 2$. 

The third model analyzed here is a two-component one which belongs to the class 
of the Villain-Lai-Das Sarma (VLDS) equation \cite{villain,laidassarma}. Again,
RD has
probability $1-p$. With probability $p$, aggregation is allowed only at valleys
or plateaus, similarly to the
RD-RSOS case. However, if aggregation is not possible at the column of
incidence, then the incident particle migrates to the nearest column in which
that condition is satisfied, and is irreversibly attached there.
When $p=1$, we obtain the conserved restricted solid-on-solid
(CRSOS) model of Kim et al \cite{kim}, where heights differences between
neighboring columns do not exceed $1$. However, with $p<1$, differences of
column heights larger than one appear due to the RD component. The competitive
model will be called RD-CRSOS.

The original CRSOS model, as well as the RD-CRSOS, are
represented in the continuum limit by the VLDS equation (also called nonlinear
molecular-beam epitaxy equation) \cite{huang,park}
\begin{equation}
{{\partial h}\over{\partial t}} = \nu_4{\nabla}^4 h +
\lambda_{4} {\nabla}^2 {\left( \nabla h\right) }^2 + \eta (\vec{x},t) ,
\label{vlds}
\end{equation}
where $\nu_4$ and $\lambda_{4}$ are constants. The best known estimates of
scaling exponents for the VLDS class were obtained from extensions of the CRSOS
model; in $d=1$, they are $\alpha =0.93$ and $z=2.88$ \cite{reiscrsos}, which
gives $\beta = 0.323$.

The RD-CRSOS model was simulated in one-dimensional
lattices with the same values of $p$ and $L$ of the RD-RSOS model. In order to
estimate the amplitudes $A$ and $C$ (Eqs. \ref{fv} and
\ref{scalingwgr}), we used the above estimates of $\alpha$ and $\beta$. In Fig.
5c we show
$\delta_p$ versus $p^2$, which suggests
$\delta\approx 1$, although the oscillations in the effective exponents do not
allow a reliable extrapolation to $p\to 0$. In Fig. 5d we show $\gamma_p$
versus $p^2$, which suggests $\gamma\approx 1/3$. These values also lead to
$y\approx 2$.

Thus, we have shown that a model in the VLDS class also has the crossover
exponents of the previous solid-on-solid models in the EW and the KPZ classes,
differing only from the value of the RD-BD model.

\section{Scaling theory for the crossover from random to correlated growth}

First we discuss the universal features of this crossover in lattice models in
any substrate dimension $d$, despite the illustrations are all given in $d=1$
for simplicity (e. g. in Fig. 2). We consider here models without additional
crossovers between different growth dynamics.

It is reasonable to assume that $t_0$ is the time in which
the roughness of random growth (Eq. \ref{wrandom}) matches the growing roughness
of the correlated process, Eq. (\ref{scalingwgr}) (see Fig. 1). This gives
\begin{equation}
C\sim {t_0}^{1/2-\beta} .
\label{Ctx}
\end{equation}
Now $t_0$ is the characteristic time for the onset of
correlations among neighboring columns, which otherwise randomly grow. On the
other hand, in a pure correlated model, the time $\delta t=1$ for deposition of
one monolayer is enough to produce such correlations. Consequently, in models
with the random to correlated crossover, we expect that
all characteristic times are scaled by a factor $t_0$.
Since the amplitude of the saturation time $t_\times$ is
$B\sim 1$ for correlated models without additional crossover (BD, RDSR,
RSOS, CRSOS, among others), we expect that
\begin{equation}
B\sim t_0 
\label{Btx}
\end{equation}
when the crossover from random growth is present.

Now substituting the amplitude $B$ from Eq. (\ref{Btx}) in the Family-Vicsek
relation (\ref{fv}) and considering that $f(x)\sim t^\beta$ in the growth
regime ($t_0\ll t\ll t_\times$), we obtain $W\sim \frac{A}{{t_0}^\beta}
t^\beta$. Comparison with Eqs. (\ref{scalingwgr}) and (\ref{Ctx}) immediately
leads to
\begin{equation}
A\sim {t_0}^{1/2} .
\label{Atx}
\end{equation}
However, this relation may be obtained from different but consistent arguments,
as follows. During time $t_0$ the neighboring columns randomly grow, thus the
local roughness $W_l$ is of the order of the RD roughness ${t_0}^{1/2}$. $W_l$
represents height fluctuations within narrow windows, whose sizes are of the
order of one lattice unit, in a large lattice and at times long enough for
significant correlations to appear inside those windows. On the other hand,
when the global roughness $W$ attains saturation, the whole system is also
highly correlated. Thus, if the total system size is $L\sim 1$ (i.e. system size
is small but not equal to $1$), this system
is a narrow window, and we expect $W_{sat}\sim W_l$. Since 
$W_{sat}\approx A$ in this case (Eq. \ref{fv} with $L=1$), we obtain Eq.
(\ref{Atx}). Certainly this argument does not apply to systems with anomalous
scaling, where local and global roughness have different scaling properties,
but that is not the case of the lattice models analyzed here or in related
works.

Now we consider the particular properties of the lattice models, focusing on
their small length-scale features. Our arguments are similar to those of
Braustein and Lam \cite{lam} for the RD-BD and the RD-RDSR models, but
here we will emphasize the independence of the results on the universality
class of the process.
In all cases, correlated growth attempts have probability $p$, while
uncorrelated growth takes place with probability $1-p$. Thus, in a given column,
the mean time interval between two successive depositions that buildup the
correlations along the substrate is $\tau = 1/p$.

Since BD involves lateral aggregation, a single deposition attempt following
this model rules introduces height correlations between that column and the
neighboring ones. This is illustrated by the deposition in column 3 of Fig.
2a: the large height difference from column 4 is immediately suppressed by
lateral aggregation. Consequently, in the RD-BD model, correlations between
neighboring columns are buildup within an average time interval $\tau$.
Consequently, we expect $t_0\sim \tau$ in any spatial dimension.
Using Eqs. (\ref{defdelta}), (\ref{defy}), (\ref{Btx}) and (\ref{Atx}),
we obtain the exponents $\delta=1/2$ and $y=1$ for the RD-BD model,
while Eqs. (\ref{defgamma}) and (\ref{Ctx})
provide exponents $\gamma$ dependent on $\beta$ and,
consequently, dependent on the substrate dimension.

Now we consider the RD-RSOS model as a typical example of solid-on-solid model.
In this case, a single RSOS attempt is not sufficient to balance the random
growth of neighboring columns. For instance, consider the aggregation rejection
in column $3$ of Fig.2b: within the time interval
$\tau$, a single particle is expected to be deposited at
columns $2$ and $4$, but these events do not suppress the
large height difference from column $3$. Thus, within a time interval of order
$\tau$, significant local correlations are not generated.
Instead, in order to the rejection mechanism to
balance the random growth locally, it is necessary that the number of
rejected attempts at a given column be of the same order of the height
difference between the neighbors. While the number of
RSOS attempts at a given column during time $t$ increases as $tp$, the
local height difference in random growth increases as $t^{1/2}$. Matching these
values we obtain the crossover time $t_0\sim p^{-2}$ for the RD-RSOS
model. This result gives $\delta=1$ and $y=2$ in all dimensions.
The exponent $\gamma$ depends on the substrate dimension due to its
dependence on the scaling exponent $\beta$ (Eqs. \ref{defgamma} and
\ref{Ctx}).

The same arguments apply to the crossover from random to EW
scaling observed in the RD-RDSR model~\cite{albano1,albano3}, since a single
diffusion event to a lower height does not suppress a large height
difference immediately. In the RD-CRSOS model, there is a combination of
rejection of aggregation at high columns and diffusion to plateaus or valleys,
but this combination also does not introduce significant height correlations
immediately. Consequently, $t_0\sim p^{-2}$ for these models, which lead to
$\delta=1$ and $y=2$.

The features of the generalized RSOS model are explained by
extending the arguments of Ref. \protect\cite{chien}.
Chien et al~\cite{chien} argued that the KPZ growth takes
place when the randomly growing roughness (Eq. \ref{wrandom}) is of order $S$,
which gives $t_0\sim S^2$. This is the same form of the previous solid-on-solid
models, with $S$ interpreted as a reciprocal probability of rejecting the
aggregation. Thus, $\delta=1$ and $y=2$ for the generalized RSOS model.

All the above results are in full agreement with simulation data shown in Sec.
II. Although our simulation results (Sec. II) were limited to
one-dimensional systems, Horowitz and Albano \cite{albano3} presented
simulation results for the RD-BD and the RD-RDSR models in $d=1$, $d=2$ and
$d=3$, an in all cases they agree with our predictions. In fact, no reference
to a particular system dimension was done in the above arguments. Instead, they
were only based on random growth properties (Eq. \ref{wrandom}), the
Family-Vicsek
relation (Eq. \ref{fv}) and the hypothesis of the existence of a crossover from
random to correlated growth.

Our analysis clearly separated solid-on-solid models with single particle
aggregation attempts and models with lateral aggregation. Other limited mobility
growth models may be classified in one of these two groups by inspection of the
microscopic aggregation rules of the correlated process. Since the
exponents $\delta=1/2$ and $y=1$ are found in models with some type of
lateral growth and, consequently, excess velocity, they are typical of
models in the KPZ class. However, models in $\delta=1$ and $y=2$ may be found
in any universality class of interface growth, including KPZ. Moreover, although
most systems previously studied are two-component models, the application to
the generalized RSOS model shows that the crossover exponents and the above
scaling approach may be extended to other systems presenting crossover from
random to correlated growth.

Now we consider the particular case of KPZ systems in $d=1$, where
relations between scaling amplitudes and the
coefficients of the growth equation are known \cite{af}:
\begin{equation}
A \sim \nu^{-1/2} , 
\label{scalingA}
\end{equation}
\begin{equation}
B \sim \nu^{1/2} {|\lambda |}^{-1} ,
\label{scalingB}
\end{equation}
and
\begin{equation}
C \sim \nu^{-2/3} {|\lambda |}^{1/3} .
\label{scalingC}
\end{equation}
In systems with a crossover from random to correlated growth, $\lambda$ and
$\nu$ may be arbitrarily small (but non-zero), while the noise amplitude $D$ is
finite, thus the dependence on $D$ is
omitted in Eqs.
(\ref{scalingA}), (\ref{scalingB}) and (\ref{scalingC}).
From Eqs. (\ref{Atx}) and (\ref{scalingA}), we obtain
\begin{equation}
t_0\sim\nu^{-1} ,
\label{txd1}
\end{equation}
and using Eqs. (\ref{Btx}) and (\ref{scalingB}) we obtain
\begin{equation}
|\lambda | \sim {\nu}^{3/2} 
\label{univrdkpz}
\end{equation}
in the crossover region of small $\nu$ and small $|\lambda |$.

These relations may also be obtained from the condition
that the random-KPZ crossover takes place at the same time of the EW-KPZ
crossover, $t_c\sim {\left( \frac{E}{C}\right)}^{12} \sim \nu^5 \lambda^{-4}$
\cite{chamereis,nt,ggg,forrest}. In other
words, after leaving the random growth regime, those systems do not show an
additional crossover from EW to KPZ because the linear and nonlinear effects
simultaneously turn up.

Eqs. (\ref{txd1}) and (\ref{univrdkpz}) provide relations between $\nu$,
$|\lambda |$ and the model parameters $p$ or $S$. $\nu\sim p$ and $|\lambda
|\sim p^{3/2}$ are obtained as particular relations for the RD-BD growth, but
cannot be viewed as universal relations for the random to KPZ crossover in
$d=1$. It contrasts to what could be naively believed from some works on
competitive models with random to correlated crossover
\cite{albano2,albano3,lam}. Instead, a large variety of KPZ processes, such as
the RD-RSOS and generalized RSOS models, follow the relations $\nu\sim p^2$
and $|\lambda |\sim p^3$ in the crossover region in $d=1$.
At this point, it is important to recall that previous numerical work on the
RD-BD model did not calculate $\nu$ and $\lambda$ directly from
the growth velocities and interface shapes \cite{albano2,albano3}. Instead,
their relations with the probability $p$ were obtained from the
scaling relations (\ref{scalingA}), (\ref{scalingB}), and (\ref{scalingC}),
similarly to what was done in other competitive models \cite{chamereis} and in
the present work.

\section{Conclusion}

We studied limited mobility growth models with crossover from random to
correlated growth.
Universal relations between the crossover time
$t_0$ and the amplitudes of saturation roughness and saturation time
were obtained from random deposition properties and Family-Vicsek scaling, for
any spatial dimension. The lattice models with that crossover can be separated
in two groups: the first one with lateral aggregation, such as ballistic
deposition, in which correlations spread faster, and the second one of the
solid-on-solid models with single particle aggregation attempts, which require
much longer times for the correlated aggregation to balance the random growth.
While $t_0\sim p$ in the first group, where $p$ is the small probability of
the correlated mechanism to work, in the second class we showed
that $t_0\sim p^2$. These relations are independent of the universality class
of the process: although the first group is expected to include only KPZ
processes, due to the presence of lateral growth, several KPZ models are also
found in the second group. All these features are confirmed by simulations of
lattice models in three different universality classes and different substrate
dimensions, partly
obtained from other authors work.  In $d=1$, relations between the probability
$p$ and the coefficients of the KPZ equation can be obtained for both groups
of models.

Many natural and artificial processes involve the competition of different
growth dynamics, and their analysis may
eventually be improved by extensions of our scaling arguments. In the
case of KPZ scaling, although the connection between scaling amplitudes and the
coefficients of the continuous equation is not trivial in $d>1$, the numerical
calculation of $\nu$ and $\lambda$ may help the search for those relations, at
least in the small $\nu$, small $\lambda$ limit.


\vfill\eject

\begin{figure}[!h]
\includegraphics[clip,width=0.80\textwidth, 
height=0.40\textheight,angle=0]{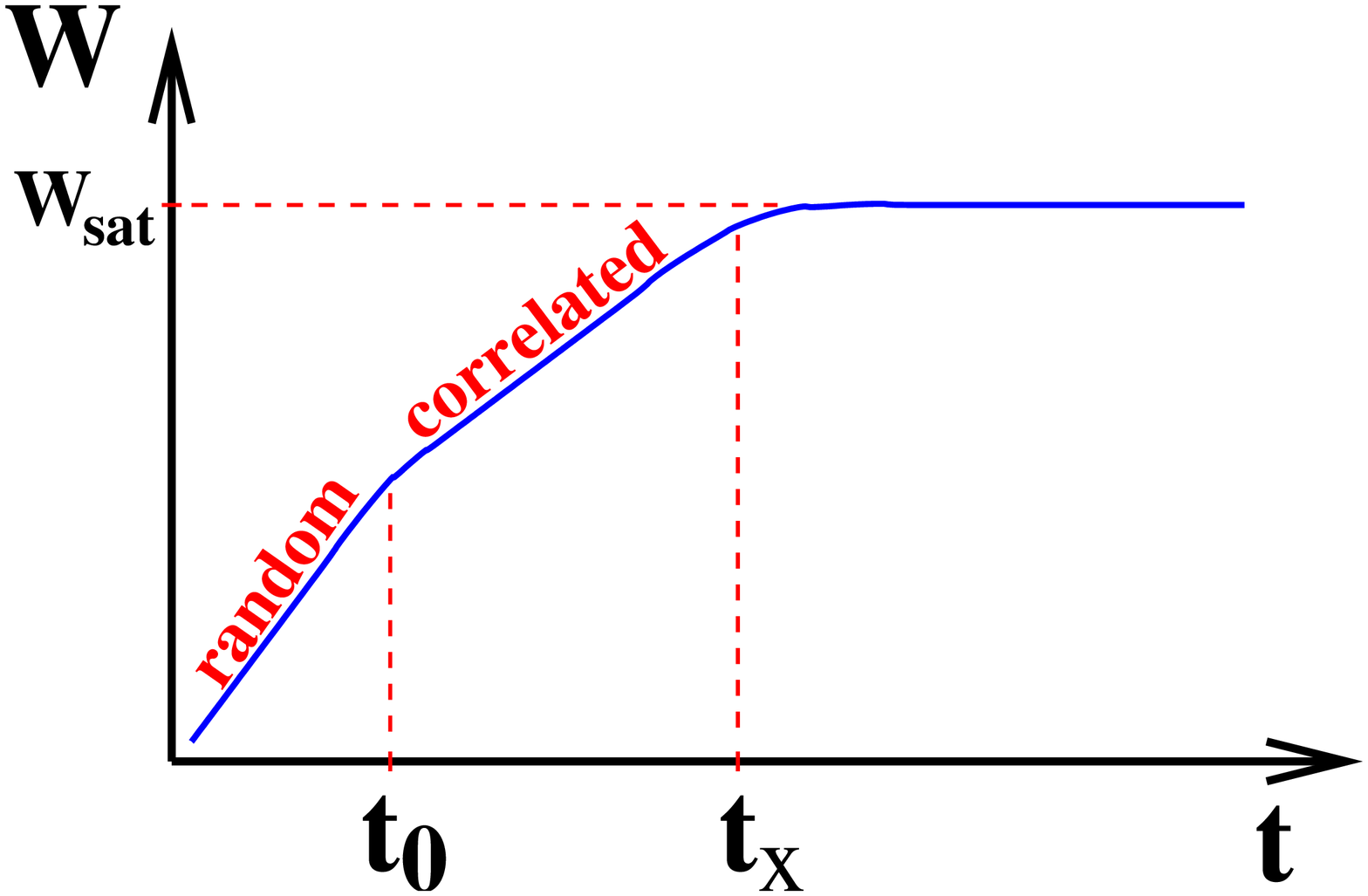}
\caption{\label{fig1} Typical time behavior of the interface width $W$ in a
system with crossover from random to correlated growth at time $t_0$, and
crossover to saturation at time $t_\times$.}
\end{figure}

\begin{figure}[!h]
\includegraphics[clip,width=0.80\textwidth, 
height=0.57\textheight,angle=0]{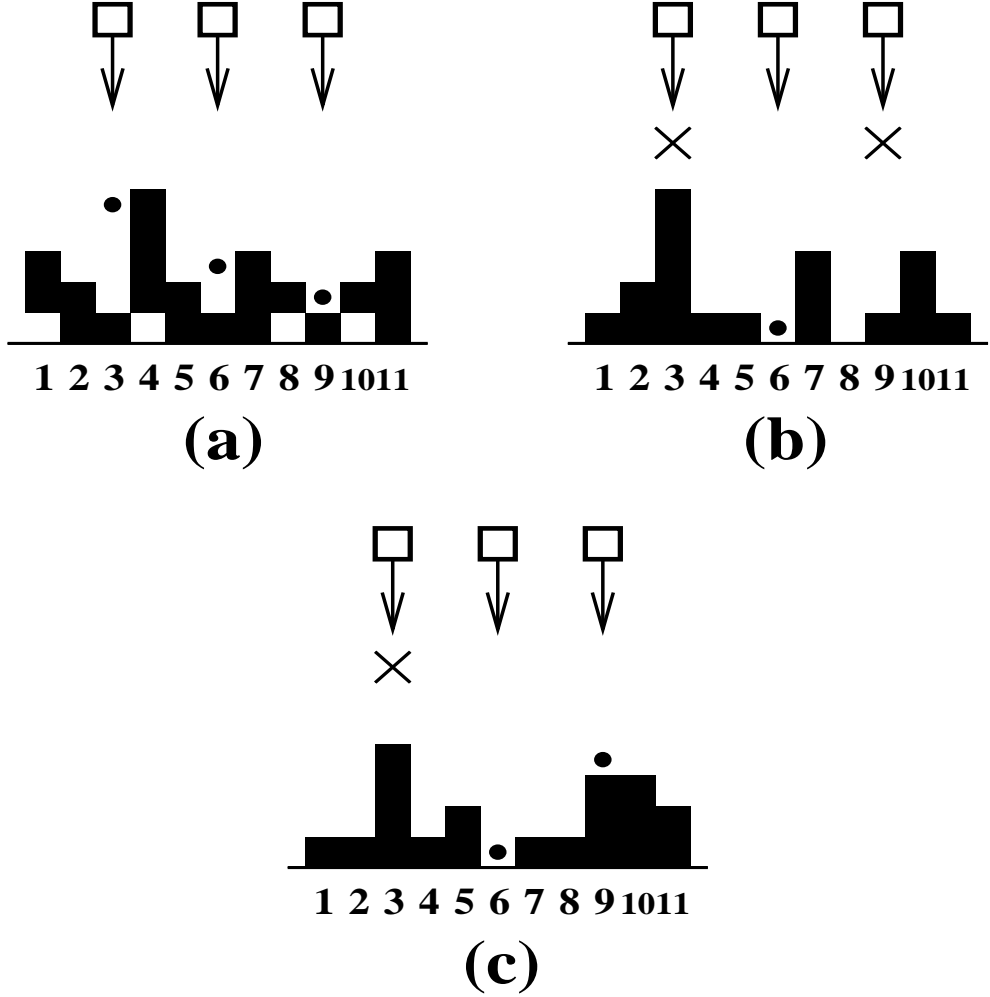}
\caption{\label{fig2} Aggregation rules of lattice models in $d=1$, in which
full squares
represent aggregated particles, open squares represent incident particles,
with the column of incidence indicated by arrows, small bullets indicate the
aggregation position of the incident particles and crosses indicate rejected
attempts of deposition. The rules of ballistic deposition are illustrated in
(a). The conditions for RSOS aggregation (probability $p$) in the competitive
model RD-RSOS are illustrated in (b). The generalized RSOS model with
$S=3$ is illustrated in (c). Column labels are indicated below the substrate
lines.}
\end{figure}

\begin{figure}[!h]
\includegraphics[clip,width=0.80\textwidth, 
height=0.57\textheight,angle=0]{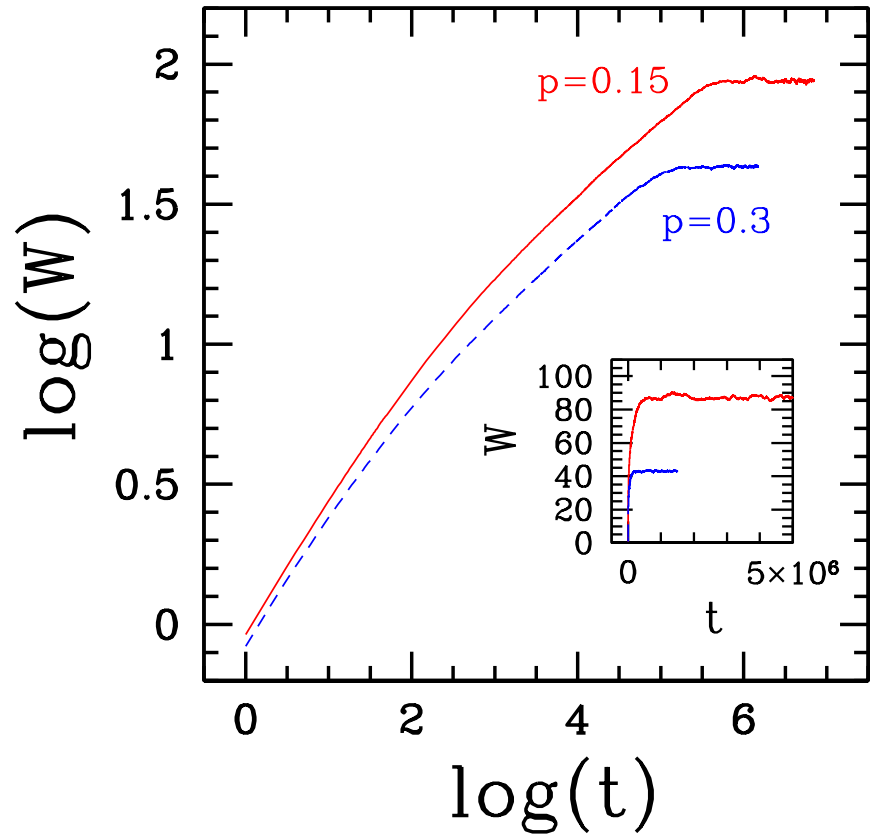}
\caption{\label{fig3} Time evolution of the roughness of the RD-RSOS model in
lattices with $L=256$, for two different probabilities $p$. The inset shows a
linear plot of the same quantities.}
\end{figure}

\begin{figure}[!h]
\includegraphics[clip,width=0.80\textwidth, 
height=0.57\textheight,angle=0]{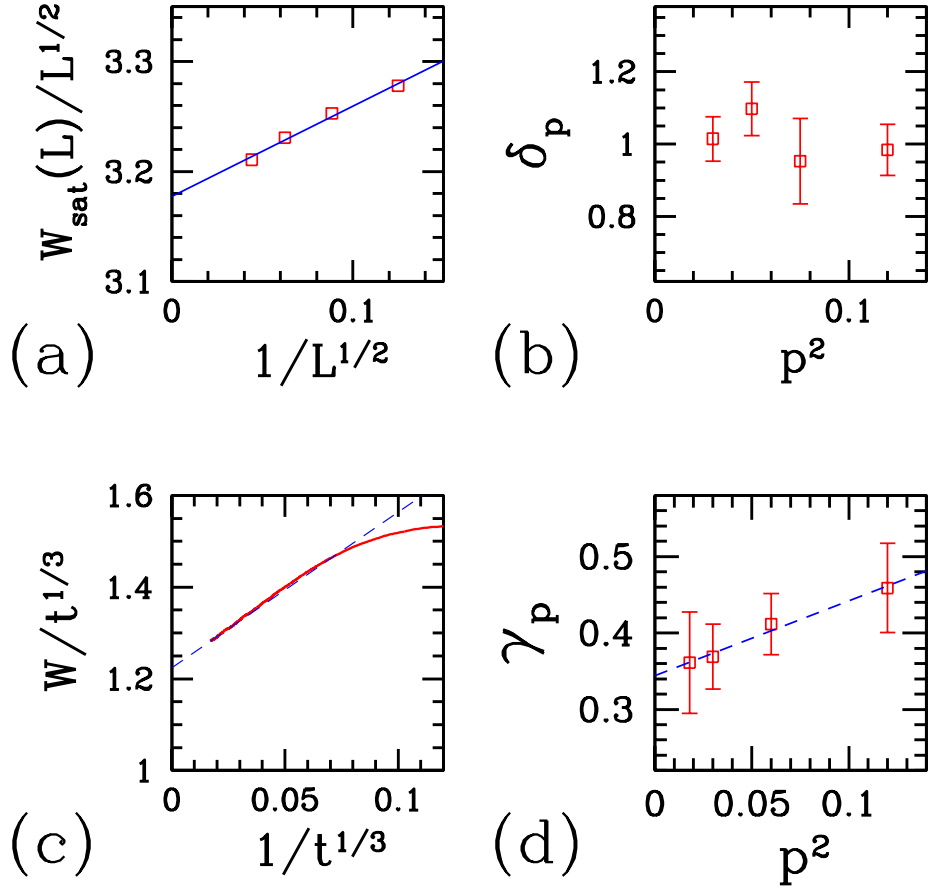}
\caption{\label{fig4} (a) Extrapolation of the ratio $W_{sat}/L^\alpha$ to
$L\to\infty$, with $\alpha=1/2$, for the RD-RSOS model in $d=1$ with $p=0.25$
and $64\leq L\leq 512$. (b) Effective exponents $\delta_p$ versus squared
probability $p^2$ for the RD-RSOS model. (c) Extrapolation of the ratio
$W/t^\beta$ to $t\to\infty$, with $\beta =1/3$, for the RD-RSOS model in $d=1$
with $p=0.2$ and $L=8192$. (d) Effective exponents $\gamma_p$ versus squared
probability $p^2$ for the RD-RSOS model, with a least squares fit of the data.}
\end{figure}

\begin{figure}[!h]
\includegraphics[clip,width=0.80\textwidth, 
height=0.57\textheight,angle=0]{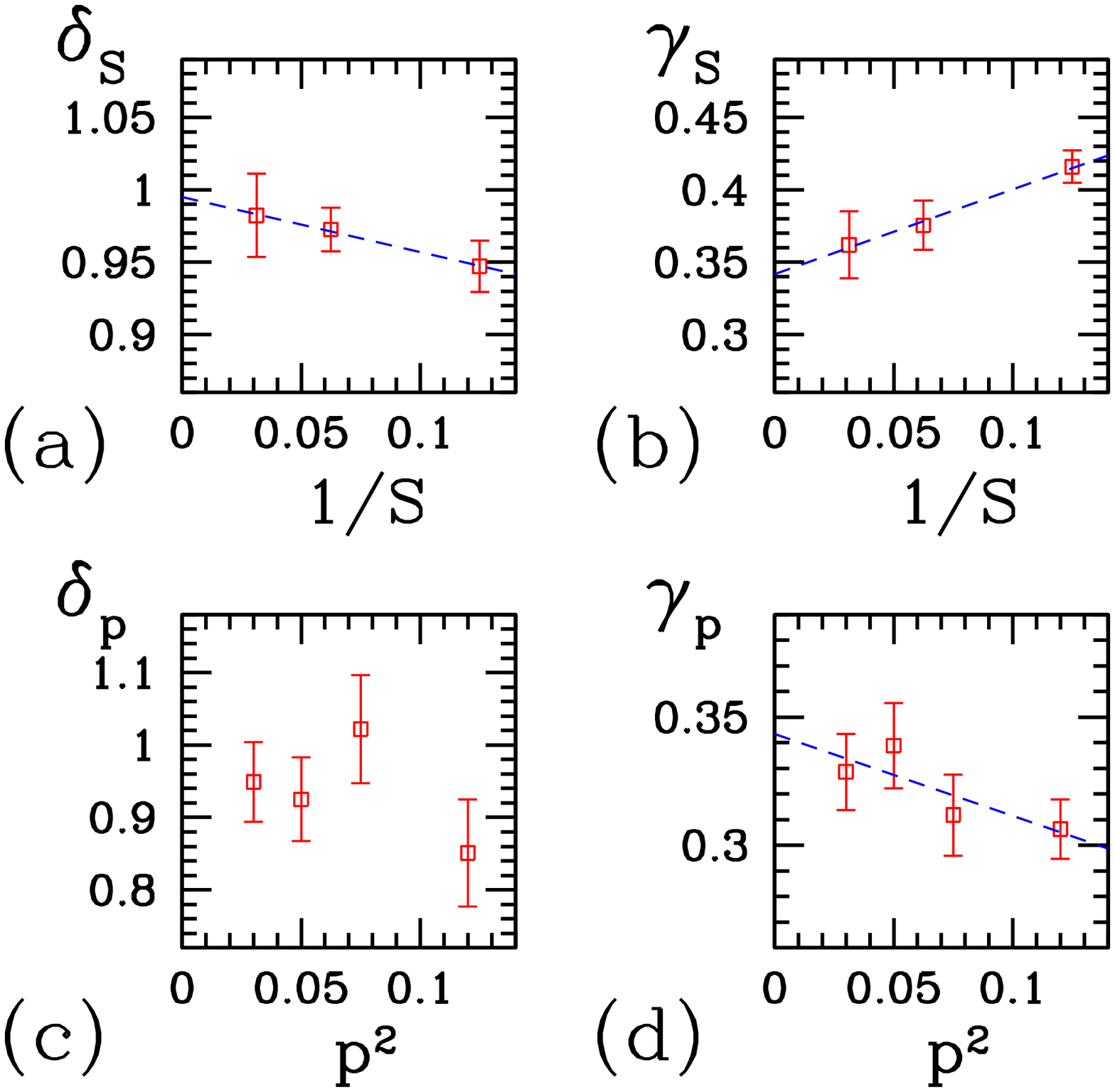}
\caption{\label{fig5} (a), (b): effective exponents $\delta_S$ and $\gamma_S$
versus $1/S$
for the generalized RSOS model; (c), (d): effective exponents $\delta_p$ and
$\gamma_p$ versus squared probability $p^2$ for the RD-CRSOS model. The dashed
lines in (a), (b) and (d) are least squares fits of the data.}
\end{figure}


\begin{references}

\bibitem{mbe}
{\it Frontiers in Surface and Interface Science}, edited by Charles B. Duke and
E. Ward Plummer (Elsevier, Amsterdam, 2002).

\bibitem{barabasi}
A.-L. Barab\'asi and H. E. Stanley, {\it Fractal Concepts in Surface Growth}
(Cambridge University Press, New York, 1995).

\bibitem{krug} J. Krug, Adv. Phys. {\bf 46}, 139 (1997).

\bibitem{cerdeira} W. Wang and H. A. Cerdeira, Phys.Rev. E {\bf 47}, 3357
(1993); H. F. El-Nashar and H. A. Cerdeira, Phys. Rev. E {\bf 61}, 6149
(2000).

\bibitem{poison1} F. D. A. Aar\~ao Reis, Phys. Rev. E {\bf 66}, 027101 (2002);
F. D. A. Aar\~ao Reis, Phys. Rev. E {\bf 68}, 041602 (2003).

\bibitem{kotrla} M. Kotrla, J. Krug and P. Smilauer, Phys. Rev. B {\bf 62},
2889 (2000).

\bibitem{albano1} C. M. Horowitz, R.A. Monetti, E.V. Albano, Phys. Rev. E
        {\bf 63}, 66132 (2001). 

\bibitem{shapir} Y. Shapir, S. Raychaudhuri, D. G. Foster, and J. Jorne, Phys.
Rev. Lett. {\bf 84}, 3029 (2000).

\bibitem{julien} Y. P. Pellegrini and R. Jullien, Phys. Rev. Lett. 
        {\bf 64} 1745 (1990); Y. P. Pellegrini and R. Jullien, Phys. Rev. A
        {\bf 43} 920 (1991).                             

\bibitem{albano2} C. M. Horowitz and E. V. Albano, J. Phys. A: Math. Gen. 
        {\bf 34}, 357 (2001).

\bibitem{albano3} C. M. Horowitz and E. V. Albano, Eur. Phys. J. B {\bf 31}, 563
(2003).

\bibitem{tales} T. J. da Silva and J. G. Moreira, Phys. Rev. E {\bf 63},
041601 (2001). 

\bibitem{chamereis} A. Chame and F. D. A. Aar\~ao Reis, Phys. Rev. E {\bf 66},
051104 (2002). 

\bibitem{lidia} D. Muraca, L. A. Braunstein, and R. C. Buceta, Phys. Rev. E
{\bf 69}, 065103(R) (2004).

\bibitem{kolakowska1}
A. Kolakowska, M. A. Novotny and P. S. Verma, Phys. Rev. E {\bf 70}, 051602
(2004).

\bibitem{lam} L. A. Braunstein and C.-H. Lam, Phys. Rev. E {\bf 72}, 026128
(2005).

\bibitem{kolakowska2}
A. Kolakowska, M. A. Novotny and P. S. Verma, cond-mat/0509668 (2005).

\bibitem{fv} F. Family and T. Vicsek, J. Phys. A {\bf 18}, L75 (1985).

\bibitem{chien} C.-C. Chien, N.-N. Pang, and W.-J. Tzeng, Phys. Rev. E {\bf
70}, 021602 (2004).

\bibitem{vold} M. J. Vold, J. Coll. Sci. {\bf 14}, 168 (1959); J. Phys.
Chem. {\bf 63}, 1608 (1959).

\bibitem{family} F. Family, J. Phys. A {\bf 19}, L441 (1986).

\bibitem{ew} S.F. Edwards and D.R. Wilkinson, Proc. R. Soc. London 
{\bf 381}, 17 (1982).

\bibitem{kpz} M. Kardar, G. Parisi and Y.-C. Zhang, Phys. Rev. Lett. 
             {\bf 56} 889 (1986).

\bibitem{marinari}
E. Marinari, A. Pagnani and G. Parisi, J. Phys. A {\bf 33} 8181 (2000).

\bibitem{kpz2d}
F. D. A. Aar\~ao Reis, Phys. Rev. E {\bf 69} 021610 (2004).

\bibitem{kk}
J. M. Kim and J. M. Kosterlitz, Phys. Rev. Lett. {\bf 62} 2289 (1989).

\bibitem{balfdaar}
F. D. A. Aar\~ao Reis, Phys. Rev. E {\bf 63} 056116 (2001);
F. D. A. Aar\~ao Reis, Physica A, to be published (2005).

\bibitem{alanissila} J. M. Kim, J. M. Kosterlitz and T. Ala-Nissila, J. Phys.
A {\bf 24}, 5569 (1991).

\bibitem{villain}
J. Villain, J. Phys. I {\bf 1} 19 (1991).

\bibitem{laidassarma}
Z.-W. Lai and S. Das Sarma, Phys. Rev. Lett. {\bf 66} 2348 (1991).

\bibitem{kim}
Y. Kim, D. K. Park and J. M. Kim, J. Phys. A: Math. Gen. {\bf 27}, L533 (1994).

\bibitem{huang}
Z.-F. Huang and B.-L. Gu, Phys. Rev. E {\bf 57}, 4480 (1998).

\bibitem{park}
S.-C. Park, D. Kim and J.-M. Park, Phys. Rev. E {\bf 65}, 015102(R) (2002).

\bibitem{reiscrsos}
F. D. A. Aar\~ao Reis, Phys. Rev. E {\bf 70}, 031607 (2004).

\bibitem{af} J.G. Amar and F. Family, Phys. Rev. A {\bf 45}, R3373 (1992).

\bibitem{nt} T. Nattermann and L.-H. Tang, Phys. Rev. A {\bf 45}, 7156 (1992).

\bibitem{ggg} B. Grossmann, H. Guo, and M. Grant, Phys. Rev. A {\bf 43}, 1727
(1991).  

\bibitem{forrest} B.M. Forrest and R. Toral, J. Stat. Physics {\bf 70}, 703
(1993).


\end{references}
\end{document}